\documentstyle[aps,preprint]{revtex}
\newcommand{\be}{\begin{equation}}
\newcommand{\ee}{\end{equation}}
\newcommand{\Be}{\begin{eqnarray}}
\newcommand{\Ee}{\end{eqnarray}}
\newcommand{\f}{\frac}
\begin{document}
\tighten
\title{Exact solutions of charged wormhole}
\author{
Sung-Won Kim\footnote{E-mail address: sungwon@mm.ewha.ac.kr
}
and Hyunjoo Lee\footnote{E-mail address: hjlee@mm.ewha.ac.kr
}}

\address{Department of Science Education
\\
Ewha Womans University
\\
Seoul 120-750, Korea}

\maketitle

\begin{abstract}

In this paper, the backreaction to the traversable Lorentzian
wormhole spacetime by the scalar field or electric charge is
considered to find the exact solutions. The charges play the role
of the additional matter to the static wormhole which is already
constructed by the exotic matter. The stability conditions for
the wormhole with scalar field and electric charge are found from
the positiveness and flareness for the wormhole shape function.

\end{abstract}
\vskip 1cm

\newpage

\section{Introduction}

To make a Lorentzian wormhole traversable, one has usually  used
an exotic matter which violates the well-known energy conditions
according to the need of the geometrical
structure\cite{MT88,MTY88}. Even though various efforts were
tried to avoid from using the exotic matter, all they were in
vain. Among the models, a wormhole in the inflating cosmological
model still required the exotic matter to be traversable and to
maintain its shape\cite{R93}. It was known that the vacuum energy
of the inflating wormhole does not change the sign of the
exoticity function. However, a traversable wormhole in the
Friedmann-Robertson-Walker(FRW) cosmological model did not
necessarily require the exotic matter at the very early
times\cite{K96}. The result meant that there was an exotic period
in the very early universe. As another example, when the wormhole
was modified by the generally coupled, massive, classical scalar
field without any geometrical modification, the exotic property
was compensated with the scalar field\cite{KK98}. For these or
other reasons, the problem about the maintenance of the wormhole
by other fields relating with the exotic property has also been
interesting to us.

There are two ways to generalize or modify the Lorentzian
traversable wormhole spacetime. (From now on `wormhole' will be
simply used as the meaning of the `Lorentzian traversable
wormhole' unless there is a confusion.) One way is the
generalization of the wormhole by alternative theory, for example,
Brans-Dicke theory\cite{AC95}. The other way is the
generalization by adding the extra matter.

As the examples of the  second generalization, the exact
solutions of the wormhole with classical, minimally-coupled,
massless scalar field and electric charge are discussed in this
paper. The backreactions of the scalar field and the electric
charge on wormhole spacetime are also found to see the
stabilities of the wormhole. We found the stability conditions
for the wormhole with scalar field and electric charge from the
positiveness and flareness for the wormhole shape function.

The similar works about the scalar field effect on wormhole have
been done by several authors. Taylor and Hiscock\cite{TH97}
examined whether the stress-energy of quantized fields in fact
will have the appropriate form to support a wormhole geometry.
They do not attempt to solve the semiclassical Einstein equations.
They found that the stress-energy tensor of the quantized scalar
field is not even qualitatively of the correct form to support the
wormhole. To maintain a wormhole classically, Vollick\cite{V97}
found the effect of the coupling a scalar field to matter which
satisfies the weak energy condition. Kim and Kim\cite{KK98} found
the solutions of the wormhole with the various scalar fields which
shares the role of the exotic matter with other matters. In this
case, the stability of the wormhole by the scalar field was not
discussed.

\section{Generalization of wormhole by extra matter}

Let the Einstein equation for the simplest normal (usual)
wormhole spacetime be \be G^{(0)}_{\mu\nu} = 8\pi
T^{(0)}_{\mu\nu}. \label{eq:ein} \ee The left hand side,
$G^{(0)}_{\mu\nu}$, is the wormhole geometry and the right hand
side, $T^{(0)}_{\mu\nu}$, is the exotic matter violating the
known energy conditions. The exotic matter is basically required
to construct the wormhole.

If the additional matter $T^{(1)}_{\mu\nu}$ is added to the right
hand side and the corresponding backreaction $G^{(1)}_{\mu\nu}$
is added to the left hand side, the Einstein equation becomes \be
G^{(0)}_{\mu\nu}+G^{(1)}_{\mu\nu}=8\pi
[T^{(0)}_{\mu\nu}+T^{(1)}_{\mu\nu} ]. \label{eq:ein3} \ee The sum
of the matters of the right hand side satisfies the conservation
law naturally. However, there is no guarantee on the exoticity of
the total matter. When the effect on the geometry of wormhole,
$G^{(1)}_{\mu\nu}$, is large enough to dominate any structure,
the additional matter might prevent from sustaining the wormhole.

Apparently, the structure of FRW cosmological model with
wormhole\cite{K96} is similar to Eq.~(\ref{eq:ein3}), but not
exactly same. In this model,  $G^{(1)}_{\mu\nu}$ is the
cosmological term which plays the role of the background
spacetime and $T^{(1)}_{\mu\nu}$ is the cosmic part of the
matter. They are not the additional term like above.

There might be an interaction term $T^{\rm (int)}_{\mu\nu}$
between $T^{(0)}_{\mu\nu}$ and $T^{(1)}_{\mu\nu}$ such as
Ref.\cite{V97}. The model is the maintaining wormhole by the
coupling a scalar field to matter that is not exotic, but the
coupling is exotic. There also might be the interaction term
$G^{\rm (int)}_{\mu\nu}$ in geometry. In some case, $G^{\rm
(int)}_{\mu\nu}$ may be joined in Eq.~(\ref{eq:ein3}) without
$T^{\rm (int)}_{\mu\nu}$, so that the equation can have the form
as $G^{(0)}_{\mu\nu}+G^{(1)}_{\mu\nu}+G^{\rm (int)}_{\mu\nu}
=8\pi (T^{(0)}_{\mu\nu}+T^{(1)}_{\mu\nu})$. This example is the
electrically charged wormhole as we will see later.

\section{Exact solutions of wormhole}

\subsection{Static wormhole}

The spacetime of the static wormhole without any charge is given
by \be ds^2 = -e^{2\Lambda(r)}dt^2 + \f{dr^2}{1-b(r)/r} + r^2
(d\theta^2+\sin^2\theta d\phi^2). \label{eq:metric} \ee The
arbitrary functions $\Lambda(r)$ and $b(r)$ are defined as the
lapse and wormhole shape functions, respectively.  The shape of
the wormhole is determined by $b(r)$. There can be two
requirements about the wormhole function $b(r)$ for the wormhole
in order to be maintained. They are positiveness and flareness
conditions. As $r \rightarrow \infty$, $b(r)$ approaches $2M$
which is defined as the mass of wormhole\cite{V95}. Therefore
$b(r)$ should be defined as the positive function. And the
condition $r > b(r)$ which means the existence of the minimum
radius also support the positiveness of $b(r)$. Another condition
comes from the flare-out condition of the shape of the wormhole.
When the proper distance $l \in (-\infty, +\infty) $ is defined as
$dl=dr/(1-b/r)$, the condition should be \be \f{d^2l}{dr^2} > 0
\qquad \mbox{or} \qquad \f{b-b'r}{b^2}
> 0. \ee

The Einstein equation Eq.~(\ref{eq:ein}) for the metric
Eq.~(\ref{eq:metric}) is given as \Be &&\f{b'}{8\pi r^2} =
\rho^{(0)},
\label{eq:worm11} \\
&&\f{b}{8\pi r^3} - \f{1}{4\pi} \left( 1- \f{b}{r} \right) \f{\Lambda'}{r}= \tau^{(0)},
\label{eq:worm12} \\
&& \f{1}{8\pi}\left( 1- \f{b}{r} \right) \left( \Lambda'' -
\f{b'r-b}{2r(r-b)}\Lambda' + \Lambda'^2 + \f{\Lambda'}{r}-
\f{b'r-b}{2r^2(r-b)} \right) = P^{(0)}, \label{eq:worm13} \Ee
where and hereafter a prime denotes the differentiation with
respect to $r$. Assuming a spherically symmetric spacetime, one
finds the components of $T_{\hat{\mu}\hat{\nu}}^{(0)} $ in
orthonormal coordinates \be T^{(0)}_{\hat{t}\hat{t}} =
\rho^{(0)}(r), ~~T^{(0)}_{\hat{r}\hat{r}} = -\tau^{(0)}(r),
~~T^{(0)}_{\hat{\theta}\hat{\theta}} = P^{(0)}(r),
\label{eq:matter} \ee where $\rho^{(0)}(r), \tau^{(0)}(r)$ and
$P^{(0)}(r)$ are  the mass energy density, radial tension per unit
area, and lateral pressure, respectively, as measured by an
observer at a fixed $r, \theta, \phi$.

\subsection{Wormhole with scalar field}

Consider the simplest case of a static Lorentzian wormhole with a
minimally-coupled massless scalar field. The additional matter
Lagrangian due to the scalar field is given by \be {\cal L} =
\f{1}{2}\sqrt{-g}g^{\mu\nu}\varphi_{;\mu}\varphi_{;\nu}
\label{eq:lag} \ee and the equation of motion for $\varphi$ by \be
\Box\varphi=0. \label{eq:wave} \ee The stress-energy tensor for
$\varphi$ is obtained from Eq. (\ref{eq:lag}) as \be
T_{\mu\nu}^{(1)} = \varphi_{;\mu}\varphi_{;\nu}
-\f{1}{2}g_{\mu\nu}g^{\rho\sigma}\varphi_{;\rho}\varphi_{;\sigma}
. \label{eq:energy} \ee Since not only the scalar field $\varphi$
but also the matter $T_{\mu\nu}$ are assumed to depend only on $r$
like the static wormhole geometry, the radiation by this scalar
field is not introduced here. The components of $T_{\mu\nu}^{(1)}
$ in the static wormhole metric with $\Lambda = 0$ have the form
\Be T_{tt}^{(1)}&=&\f{1}{2}\left(1-\f{b}{r}\right)\varphi'^2,
\label{eq:scem1}\\
T_{rr}^{(1)}&=&\f{1}{2}\varphi'^2, \label{eq:scem2}\\
T_{\theta\theta}^{(1)}&=&-\f{1}{2}r^2\left(1-\f{b}{r}\right)
\varphi'^2, \label{eq:scem3}\\
T_{\phi\phi}^{(1)}&=&
-\f{1}{2}r^2\left(1-\f{b}{r}\right)\varphi'^2\sin^2\theta.
\label{eq:scem4}
\Ee
 In this spacetime,
the field equation Eq.~(\ref{eq:wave}) of $\varphi$ becomes
\be
\f{\varphi''}{\varphi'}+\f{1}{2}\f{(1-b/r)'}{(1-b/r)}+\f{2}{r} =
0 \quad \quad \mbox{or} \quad \quad
r^4\varphi'^2\left(1-\f{b}{r}\right) = \mbox{const}.
\label{eq:phi}
\ee
The integration constant will be represented as
$\f{\alpha}{4\pi}$ for some positive $\alpha$.

If we add $G^{(1)}_{\mu\nu}$ as an additional geometry to
$G^{(0)}_{\mu\nu}$, the effect by scalar field, the Einstein
equation Eq.~(\ref{eq:ein3}) for $\Lambda=0$ is changed from
Eq.~(\ref{eq:worm11})-(\ref{eq:worm13}) into
\Be
\f{b'}{8\pi r^2}
+ \f{1}{8\pi}\f{\alpha}{r^4} &=& \rho^{(0)} + \f{1}{2}\varphi'^2
\left( 1 - \f{b}{r} \right),
\label{eq:worm21} \\
\f{b}{8\pi r^3} - \f{1}{8\pi}\f{\alpha}{r^4} &=& \tau^{(0)} - \f{1}{2}\varphi'^2
\left( 1 - \f{b}{r} \right),
\label{eq:worm22} \\
\f{b-b'r}{16\pi r^3} - \f{1}{8\pi}\f{\alpha}{r^4} &=& P^{(0)} -
\f{1}{2}\varphi'^2 \left( 1 - \f{b}{r} \right), \label{eq:worm23}
\Ee when the interaction between the matter and additional scalar
fields is neglected. The term  $\alpha/r^4$ is added to the left
hand side, because the field equation Eq.~(\ref{eq:phi}) for
$\varphi$ shows that  $\varphi'^2 \left( 1 - \f{b}{r} \right)
\propto r^{-4}$.

If we put $b_{\rm eff} = b - \alpha/r$ instead of $b$ and $T^{\rm
eff}_{\mu\nu} = T^{(0)}_{\mu\nu}+T^{(1)}_{\mu\nu}$ instead of
$T^{(0)}_{\mu\nu}$ into Eq.~(\ref{eq:worm11})-(\ref{eq:worm13})
with $\Lambda=0$, then the effective equations will satisfy
self-consistently and has the form of
Eq.~(\ref{eq:worm21})-(\ref{eq:worm23}). Thus the effect by the
scalar field on the wormhole is simply represented as the change
of the wormhole function $b$ into $(b-\alpha/r)$ without any
interaction term in the left hand side. Since $b$ is proportional
to $r^{1/(1+2\beta)}$, with the proper parameter $\beta$ of
equation of state\cite{K96}, the shape of the effective wormhole
will vary with the value of parameters $\beta$ and $\alpha$ via
the additional factor $-\alpha/r$. While $\beta$ is given as the
equation of state by the choice of the appropriate matter,
$\alpha$ depends on the changing rate of the scalar field
$\varphi$. Since $\alpha$ plays the role of the scalar charge,
the metric of the wormhole spacetime with the scalar field should
be \be ds^2 = -dt^2 + \left( 1 - \f{b(r)}{r} + \f{\alpha}{r^2}
\right)^{-1} dr^2 + r^2 ( d\theta^2 + \sin^2\theta d\phi^2 )
\label{eq:solsca} \ee in case of $\Lambda=0$. No horizon occurs
in this spacetime, since the components are always positive.

For the dimensional reasons, the wormhole function has the form as
\be
b = b_0^{\f{2\beta}{2\beta+1}}r^{\f{1}{2\beta+1}},
\ee
where
$\beta$ should be less than $-\f{1}{2}$ so that the exponent of
$r$ can be negative to satisfy the flareness condition.

The positiveness and flareness conditions for the effective
wormhole can be written as
\Be
b_{\rm eff} &>& 0, \label{eq:po1}\\
\f{b_{\rm eff}-b_{\rm eff}'r}{b_{\rm eff}^2} &>& 0. \label{eq:fl1}
\Ee
If we rewrite these in terms of $b(r)$,
\Be
b - \f{\alpha}{r} &>& 0 , \label{eq:po2} \\
\gamma b - \f{\alpha}{r} &>& 0,
\label{eq:fl2}
\Ee
where $\gamma = \f{\beta}{2\beta+1}$.

When $-1<\beta<-\f{1}{2}$, the flareness condition is included in
the positiveness condition, since $\gamma > 1$. That is, if only
$b>\f{\alpha}{r}$, the flareness condition Eq.~(\ref{eq:fl2}) is
satisfied automatically. In this case, the power  of $b(r)$ is
less than $-1$ and the function $b(r)$ vanishes more quickly than
the second term, in the far region. Thus it gives the negative
region for $b_{\rm eff}$ at large $r$, even though it has the
positive regions near throat when $b^2_0>\alpha$. At throat, the
effective wormhole shape function becomes $ \f{1}{b_0}(b^2_0 -
\alpha)$, which shows that the size of neck is reduced by the
scalar field. The region of the positive $b_{\rm eff}$ is
$b_0<r<r_0$, where $r_0 =
\alpha^{(2\beta+1)/(2\beta+2)}/b_0^{\beta/(2\beta+1)}$. If $b^2_0
< \alpha$, $b_{\rm eff}$ is negative at all $r$, which is not
suitable for wormhole.

If only $\beta \le -1$ and $b^2_0 > \alpha$, the wormhole is safe,
because $b_{\rm eff}$ is positive at all $r$. When $b^2_0 <
\alpha$, there also be a region ($r < r_0 $) of negative $b$ so
that the scalar field effect will change the wormhole structure
into others, since the scalar field dominates the exotic matter.
Thus the addition of the minimally-coupled, massless scalar field
does not guarantee the structure of wormhole.

Now we shall examine the special case of this backreaction
problem, for instance, $\beta = -1$ which is $b = b_0^2 /r$. In
this case, the solution of the scalar field  is given
as\cite{KK98} \be \varphi = \varphi_0 \left[ 1 - \cos^{-1} \left(
\f{b_0}{r} \right) \right]. \ee Thus the proportional constant
$\alpha$ becomes \be \alpha = 4\pi \varphi_0^2 b_0^2 , \ee where
$b_0$ is the minimum size of the wormhole and $\varphi_0$ is the
value of $\varphi(r)$ at $r=b_0$. Therefore, $\varphi^2_0 <
\f{1}{4\pi}$ is the condition that is required for maintaining
the wormhole under the addition of the scalar field. In this
choice of $\beta = -1$, there is no $r_0$ at which the sign of
$b_{\rm eff}$ changes.

We can also apply the result to the other form of $b(r)$, which
means the exotic matter distribution in the restricted region
only, ``absurdly benign'' wormhole,\cite{MTY88}
\be b(r) = \left\{
\begin{array}{cl} b_0 [1-(r-b_0)/a_0]^2, \Phi(r) = 0, & ~~{\rm
for}~~~~b_0
\le r \le b_0 + a_0, \\
b=\Phi=0, & ~~{\rm for}~~~~ r \ge b_0+a_0
\end{array}
\right. \ee In this case, since the second term $-\alpha/r$ in the
effective shape function extends to over the region $r \ge b_0 +
a_0$, there will be a negative $b_{\rm eff}$ within this range of
$r$, which is not safe for wormhole formation.

\subsection{Wormhole with electric charge}

In the case of electric charge, we can follow the same procedures
as the case of scalar field. However, the simplification with
$\Lambda=0$ is not adequate here, because the self-consistent
solution cannot be found in this case. Thus the work on finding
extra terms in geometry will be very complicated one, compared
with the scalar field case. We rather start from assuming the
probable spacetime metric and check later whether it will be
correct or not. We can presumably set the wormhole spacetime with
electric charge $Q$ as \be ds^2 = -\left( 1 + \f{Q^2}{r^2}
\right)dt^2 + \left(1- \f{b(r)}{r} + \f{Q^2}{r^2} \right)^{-1}dr^2
+ r^2 (d\theta^2+\sin^2\theta d\phi^2). \label{eq:charge} \ee This
spacetime is the combination of MT type spherically symmetric
static wormhole and Reissner-Nordstr\"om(RN) spacetime. When
$Q=0$, the spacetime means the MT wormhole and when $b=0$, it
becomes the RN black hole with zero mass. The metric
Eq.~(\ref{eq:charge}) should be checked whether it satisfies the
Einstein's equation self-consistently or not.

The Einstein-Maxwell equation become
\Be
&&\f{b'}{r^2} + \f{Q^2}{r^4} = 8\pi(\rho^{(0)} + \rho^{(1)}),  \\
&&\f{b}{r^3} - \f{Q^2}{r^4} - 2 \left( 1 - \f{b}{r} + \f{Q^2}{r^2} \right)
\left(-\f{Q^2}{r^2(r^2+Q^2)}\right)
 = 8\pi(\tau^{(0)} +\tau^{(1)}),
\Ee
\Be
&&\left( 1- \f{b}{r} + \f{Q^2}{r^2} \right) \left[
\f{Q^2(3r^2+Q^2)}{r^2(r^2+Q^2)^2}\right.
-\left(\f{b'r-b+\f{2Q^2}{r}}{2(r^2-br+Q^2)}\right)\left(-\f{Q^2}{r(r^2+Q^2)^2}\right)
\nonumber\\
&&+\left.\left(-\f{Q^2}{r(r^2+Q^2)^2}\right)^2 - \f{Q^2}{r^2(r^2+Q^2)^2}
-\f{b'r-b+\f{2Q^2}{r}}{2r(r^2-br+Q^2)}\right]  = 8\pi(P^{(0)} + P^{(1)}),
\Ee
The matter terms are
\be
\rho^{(1)} = \tau^{(1)} = P^{(1)} = \f{Q^2}{8\pi r^4},
\ee
since $T^{(1)}_{\mu\nu}= \f{1}{4\pi}( F_{\mu\lambda}{F^\lambda}_\nu
- \f{1}{4}g_{\mu\nu}F_{\lambda\sigma}F^{\lambda\sigma})$
and the electromagnetic strength tensor $F_{\mu\nu}$ is given as
\be
F_{\mu\nu}= E(r) \left(  \begin{array}{cccc}
~0~ & ~1~ & ~0~ & ~0~ \\
~-1~ & ~0~ & ~0~ & ~0~ \\
~0~ & ~0~ & ~0~ & ~0~ \\
~0~ & ~0~ & ~0~ & ~0~ \\   \end{array} \right) \ee by the
spherical symmetry. The electric field is given as \be E =
\f{Q}{r^2}\sqrt{g_{tt}g_{rr}} \ee by Maxwell's equation. There
are the interacting coupling terms $Q\cdot b$ in geometry, even
though no interaction terms in matter part at all. The reason is
that the zero-tidal force is not assumed in this case unlike the
scalar field case.

If
\[
b~\rightarrow~b_{\rm eff} = b - \f{Q^2}{r},
\]
the effective wormhole shape function $b_{\rm eff}$ and the total
matter $T^{\rm eff}_{\mu\nu}$ self-consistently satisfy the
equations like the scalar field case. Therefore, the metric
Eq.~(\ref{eq:charge}) can be convinced as the spacetime of
wormhole with electric charge $Q$.

 When $b=b_0^{\f{2\beta}{2\beta+1}}r^{\f{1}{2\beta+1}}$
like the case of the scalar field, there will be the same sign
relations of $b_{\rm eff}$ as shown in Table~1, except that
$\alpha$ is replaced by $Q^2$. Here, $r_0$ which changes the sign
of the wormhole becomes
\[
r_0 = Q^{\f{2\beta+1}{\beta+1}}b_0^{\f{\beta}{2\beta+1}}
\]

In the special case of $\beta = -1$, $b = b_0^2 /r$, $Q^2 <
b^2_0$ is the condition that is required for maintaining the
wormhole under the addition of the electric charge. When someone
will try to prevent from formation of a wormhole with throat of 1
meter radius, he should prepare the charge over than
$3\times10^{16}$coulombs. This amount of charge is too huge to
exist as a single kind of charge.

\vskip 1cm
\begin{center}
\begin{tabular}{|c|c|c|}
\multicolumn{3}{c}{Table 1. The sign of the effective wormhole
function $b_{\rm eff}(r)$.}
\\ \hline \hline
\multicolumn{1}{|c|}{~~~Value of $\beta$~~~} &
\multicolumn{1}{c|}{~~~~~~$b^2_0 > Q^2$ or $\alpha^\dag$~~~~~~} &
\multicolumn{1}{c|}{~~~$b^2_0 < Q^2$ or $\alpha^\dag$~~~}
\\ \hline
$\beta < -1 $ & $+$ & $- +^\ddag$
\\ \hline
$\beta = -1 $ & $+$ & $-$
\\ \hline
$ -1 < \beta < -\f{1}{2} $ & $+ -^\ddag$ & $-$
\\ \hline
\end{tabular}
\end{center}
\vskip -.3cm \hskip 3cm {\small ${}^\dag$$Q^2$ in case of
electric charge and $\alpha$ in case of scalar field. }

\vskip -.4cm \hskip 2.4cm {\small ${}^\ddag$The signs change at
$r_0$.}

\section{Discussion}

Here we studied the backreactions to the static wormhole by the
scalar field and electric charge and found the self-consistent
solutions. The charges are considered as the static case only
like the wormhole, so there is no radiation by the fields. We
found the exact solutions of wormhole with extra fields such as
scalar field and electric charge. The conditions of the wormhole
shape through the requirement of positiveness and flareness were
also found.

As we see in Eq.~(\ref{eq:solsca}) and Eq.~(\ref{eq:charge}), no
horizon occurs in these wormhole spacetimes with charges, because
of the positiveness for the metric components. It means that the
addition of charge might change the wormhole but will not change
the spacetime seriously.

In this paper, the interaction between the extra field and the
original matter is neglected. If the interaction exists and it is
large, it might change the whole geometry drastically. If it is
very small, it does not seem to change the main structure of the
wormhole. The detailed discussion on these interactions will be
in separate paper.

\acknowledgements

This research was supported by the Korea Research
Foundation(KRF-99-015-DI0021).

\end{document}